\documentclass[conference]{IEEEtran}
\IEEEoverridecommandlockouts
\usepackage{cite}
\usepackage{amsmath,amssymb,amsfonts,amsthm}
\usepackage[nolist]{acronym}
\usepackage{algorithmic}
\usepackage{hyperref}
\usepackage{scalerel}
\usepackage{graphicx}
\usepackage{textcomp}
\usepackage{xcolor}
\usepackage{tikz}
\usepackage{nicefrac}
\usepackage{circuitikz}

\usepackage{booktabs}

\usepackage{multirow}

\usepackage{siunitx}

\newdimen\bpt
\def\mobile#1{\leavevmode 
   \bpt=#1bp \hbox to7\bpt{\kern1\bpt \lower1\bpt\vbox to12\bpt{}%
      \pdfliteral{q #1 0 0 #1 0 0 cm 1 j 2 w 0 0 5 10 re B 
         1 g 1 G  1 w .3 1.8 4.4 7 re B 
         1.5 w 2.5 .2 0 .1 re B .3 w 1.7 10 1.6 0 re B Q}%
      \hss}}

\usetikzlibrary{calc}
\usetikzlibrary{positioning}
\usepackage{pgfplots}
\usepgfplotslibrary{groupplots}
\usepgfplotslibrary{colorbrewer}
\pgfplotsset{compat = 1.15}

\makeatletter
\let\MYcaption\@makecaption
\makeatother

\usepackage[font=footnotesize]{subcaption}

\makeatletter
\let\@makecaption\MYcaption
\makeatother

\DeclareMathOperator*{\argmax}{arg\,max}
\DeclareMathOperator*{\argmin}{arg\,min}

\usetikzlibrary{svg.path}

\def\BibTeX{{\rm B\kern-.05em{\sc i\kern-.025em b}\kern-.08em
    T\kern-.1667em\lower.7ex\hbox{E}\kern-.125emX}}

\definecolor{orcidlogocol}{HTML}{A6CE39}
\tikzset{
  orcidlogo/.pic={
    \fill[orcidlogocol] svg{M256,128c0,70.7-57.3,128-128,128C57.3,256,0,198.7,0,128C0,57.3,57.3,0,128,0C198.7,0,256,57.3,256,128z};
    \fill[white] svg{M86.3,186.2H70.9V79.1h15.4v48.4V186.2z}
                 svg{M108.9,79.1h41.6c39.6,0,57,28.3,57,53.6c0,27.5-21.5,53.6-56.8,53.6h-41.8V79.1z M124.3,172.4h24.5c34.9,0,42.9-26.5,42.9-39.7c0-21.5-13.7-39.7-43.7-39.7h-23.7V172.4z}
                 svg{M88.7,56.8c0,5.5-4.5,10.1-10.1,10.1c-5.6,0-10.1-4.6-10.1-10.1c0-5.6,4.5-10.1,10.1-10.1C84.2,46.7,88.7,51.3,88.7,56.8z};
  }
}

\newcommand\orcidicon[1]{\href{https://orcid.org/#1}{\mbox{\scalerel*{
\begin{tikzpicture}[yscale=-1,transform shape]
\pic{orcidlogo};
\end{tikzpicture}
}{|}}}}

\begin{acronym}
 \acro{R2M}{raw 2\textsuperscript{nd} moment}
 \acro{CSI}{channel state information}
 \acro{UE}{user equipment}
 \acro{EM}{electromagnetic}
 \acro{UL}{uplink}
 \acro{BS}{base station}
 \acro{TDD}{time division duplex}
 \acro{FDD}{frequency division duplex}
 \acro{ECC}{error-correcting code}
 \acro{MLD}{maximum likelihood decoding}
 \acro{HDD}{hard decision decoding}
 \acro{IF}{intermediate frequency}
 \acro{RF}{radio frequency}
 \acro{SDD}{soft decision decoding}
 \acro{NND}{neural network decoding}
 \acro{CNN}{convolutional neural network}
 \acro{ML}{maximum likelihood}
 \acro{GPU}{graphical processing unit}
 \acro{BP}{belief propagation}
 \acro{LTE}{Long Term Evolution}
 \acro{BER}{bit error rate}
 \acro{SNR}{signal-to-noise-ratio}
 \acro{ReLU}{rectified linear unit}
 \acro{BPSK}{binary phase shift keying}
 \acro{QPSK}{quadrature phase shift keying}
 \acro{AWGN}{additive white Gaussian noise}
 \acro{MSE}{mean squared error}
 \acro{LLR}{log-likelihood ratio}
 \acro{MAP}{maximum a posteriori}
 \acro{NVE}{normalized validation error}
 \acro{BCE}{binary cross-entropy}
 \acro{CE}{cross-entropy}
 \acro{BLER}{block error rate}
 \acro{SQR}{signal-to-quantisation-noise-ratio}
 \acro{MIMO}{multiple-input multiple-output}
 \acro{mMIMO}{massive multiple-input multiple-output}
 \acro{OFDM}{orthogonal frequency division multiplex}
 \acro{RF}{radio frequency}
 \acro{LOS}{line of sight}
 \acro{NLoS}{non-line of sight}
 \acro{NMSE}{normalized mean squared error}
 \acro{CFO}{carrier frequency offset}
 \acro{SFO}{sampling frequency offset}
 \acro{IPS}{indoor positioning system}
 \acro{TRIPS}{time-reversal IPS}
 \acro{RSSI}{received signal strength indicator}
 \acro{MIMO}{multiple-input multiple-output}
 \acro{ENoB}{effective number of bits}
 \acro{AGC}{automatic gain control}
 \acro{ADC}{analog to digital converter}
 \acro{ADCs}{analog to digital converters}
 \acro{FB}{front bandpass}
 \acro{FPGA}{field programmable gate array}
 \acro{JSDM}{Joint Spatial Division and Multiplexing}
 \acro{NN}{neural network}
 \acro{IF}{intermediate frequency}
 \acro{LoS}{line-of-sight}
 \acro{NLoS}{non-line-of-sight}
 \acro{DSP}{digital signal processing}
 \acro{AFE}{analog front end}
 \acro{SQNR}{signal-to-quantisation-noise-ratio}
 \acro{SINR}{signal-to-interference-noise-ratio}
 \acro{ENoB}{effective number of bits}
 \acro{PCB}{printed circuit board}
 \acro{EVM}{error vector mangnitude}
 \acro{CDF}{cumulative distribution function}
 \acro{MRC}{maximum ratio combining}
 \acro{MRP}{maximum ratio precoding}
 \acro{MRT}{maximum ratio transmission}
 \acro{DeepL}{deep-learning}
 \acro{DL}{downlink}
 \acro{SISO}{single-input single-output}
 \acro{SGD}{stochastic gradient descent}
 \acro{CP}{cyclic prefix}
 \acro{MISO}{Multiple Input Single Output}
 \acro{LMMSE}{linear minimum mean square error}
 \acro{ZF}{zero forcing}
 \acro{USRP}{universal software radio peripheral}
 \acro{RNN}{recurrent neural network}
 \acro{GRU}{gated recurrent unit}
 \acro{LSTM}{long short-term memory}
 \acro{NTM}{neural turing machine}
 \acro{DNC}{differentiable neural computer}
 \acro{TCN}{temporal convolutional network}
 \acro{FCL}{fully connected layer}
 \acro{MANN}{memory augmented neural network}
 \acro{RNN}{recurrent neural network}
 \acro{DNN}{deep neural network}
 \acro{FIR}{finite impulse response}
 \acro{BPTT}{back-propagation through time}
 \acro{GAN}{generative adversarial network}
 \acro{ELU}{exponential linear unit}
 \acro{tanh}{hyperbolic tangent}
 \acro{BICM}{bit-interleaved coded modulation}
 \acro{OTA}{over-the-air}
 \acro{IM}{intensity modulation}
 \acro{DD}{direct detection}
 \acro{RL}{reinforcement learning}
 \acro{SDR}{software-defined radio}
 \acro{WGAN}{Wasserstein generative adversarial network}
 \acro{BMD}{bit-metric decoding}
 \acro{BMI}{bit-wise mutual information}
 \acro{LDPC}{low-density parity-check}
 \acro{IDD}{iterative demapping and decoding}
 \acro{JSD}{Jensen-Shannon divergence}
 \acro{MMSE}{minimum mean square error}
 \acro{FFT}{fast Fourier transform}
 \acro{DFT}{discrete Fourier transform}
 \acro{IFFT}{inverse fast Fourier transform}
 \acro{QAM}{quadrature amplitude modulation}
 \acro{EMD}{earth mover's distance}
 \acro{TDL}{tapped delay line}
 \acro{KL}{Kullback-Leibler}
 \acro{PRACH}{physical random access channel}
 \acro{URLLC}{ultra-reliable low-latency communication}
 \acro{ANOMA}{asynchronous non-orthogonal multiple access}
 \acro{FEC}{forward error correction}
 \acro{PAPR}{peak-to-average power ratio}
 \acro{APP}{a posteriori probability}
 \acro{COTS}{commercial off-the-shelf}
 \acro{PLL}{phase locked loop}
 \acro{STO}{sampling time offset}
 \acro{SFO}{sampling frequency offset}
 \acro{CFO}{carrier frequency offset}
 \acro{CPO}{carrier phase offset}
 \acro{CSI}{channel state information}
 \acro{GNSS}{global navigation satellite system}
 \acro{ELAA}{extremely large aperture array}
 \acro{UE}{user equipment}
 \acro{DICHASUS}{\underline{Di}stributed \underline{Cha}nnel \underline{S}ounder by \underline{U}niversity of \underline{S}tuttgart}
 \acro{JCaS}{Joint Communication and Sensing}
 \acro{CT}{Continuity}
 \acro{TW}{Trustworthiness}
 \acro{KS}{Kruskal's stress}
 \acro{PCA}{principal component analysis}
 \acro{AoA}{angle of arrival}
 \acro{RD}{Rajski's distance}
 \acro{ADP}{angle-delay profile}
 \acro{MPC}{multipath component}
 \acro{CIR}{channel impulse response}
 \acro{MAE}{mean absolute error}
 \acro{MDS}{multidimensional scaling}
 \acro{t-SNE}{t-distributed stochastic neighbor embedding}
 \acro{SM}{Sammon's mapping}
 \acro{CIRA}{channel impulse response amplitude}
 \acro{CS}{cosine similarity}
 \acro{FCF}{forward charting function}
 \acro{CMD}{correlation matrix distance}
 \acro{6G}{Sixth generation}
 \acro{JEPA}{joint-embedding predictive architecture}

\end{acronym}

\definecolor{mittelblau}{RGB}{0, 126, 198}
\definecolor{violettblau}{cmyk}{0.9, 0.6, 0, 0}
\definecolor{rot}{RGB}{238, 28 35}
\definecolor{apfelgruen}{RGB}{140, 198, 62}
\definecolor{gelb}{RGB}{1, 221, 0}
\definecolor{orange}{RGB}{244, 111, 33}
\definecolor{pink}{RGB}{237, 0, 140}
\definecolor{lila}{RGB}{128, 10, 145}
\definecolor{hellgrau}{RGB}{224, 224, 224}
\definecolor{mittelgrau}{RGB}{128, 128, 128}
\definecolor{dunkelgrau}{RGB}{80,80,80}
\definecolor{anthrazit}{RGB}{19, 31, 31}

\begin{document}


\title{Channel Charting-Based Channel Prediction on Real-World Distributed Massive MIMO CSI\\
\thanks{This work is supported by the German Federal Ministry of Education and Research (BMBF) within the projects Open6GHub (grant no. 16KISK019) and KOMSENS-6G (grant no. 16KISK113).}
}

\author{\IEEEauthorblockN{Phillip Stephan\textsuperscript{\orcidicon{0009-0007-4036-668X}}, Florian Euchner\textsuperscript{\orcidicon{0000-0002-8090-1188}}, Stephan ten Brink\textsuperscript{\orcidicon{0000-0003-1502-2571}} \\}

\IEEEauthorblockA{
Institute of Telecommunications, Pfaffenwaldring 47, University of  Stuttgart, 70569 Stuttgart, Germany \\ \{stephan,euchner,tenbrink\}@inue.uni-stuttgart.de
}
}

\maketitle

\begin{abstract}
Distributed massive MIMO is considered a key advancement for improving the performance of next-generation wireless telecommunication systems.
However, its efficacy in scenarios involving user mobility is limited due to channel aging.
To address this challenge, channel prediction techniques are investigated to forecast future channel state information (CSI) based on previous estimates.
We propose a new channel prediction method based on channel charting, a self-supervised learning technique that reconstructs a physically meaningful latent representation of the radio environment using similarity relationships between CSI samples.
The concept of inertia within a channel chart allows for predictive radio resource management tasks through the latent space.
We demonstrate that channel charting can be used to predict future CSI by exploiting spatial relationships between known estimates that are embedded in the channel chart.
Our method is validated on a real-world distributed massive MIMO dataset, and compared to a Wiener predictor and the outdated CSI in terms of achievable sum rate.
\end{abstract}

\begin{IEEEkeywords}
AI, channel aging, channel charting, channel prediction, cell-free, distributed, massive MIMO
\end{IEEEkeywords}

\section{Introduction}
\ac{6G} wireless communication systems need to deploy new technologies to achieve the ambitious requirements in terms of data throughput, latency and energy consumption.
Distributed massive \ac{MIMO}, also known as cell-free massive \ac{MIMO}, is seen as a key enabler for spatial multiplexing by installing numerous spatially distributed antennas at the \ac{BS}.
To fully exploit their potential, these systems rely on the availability of timely \ac{CSI}.
Operating in \ac{TDD} mode, the estimation of \ac{UL} \ac{CSI} at the \ac{BS} is sufficient for \ac{DL} communication because of the reciprocity between \ac{UL} and \ac{DL} channel.
However, rapid changes of the wireless channel due to \ac{UE} mobility or moving objects lead to outdated \ac{CSI} estimates at the \ac{BS}.
This phenomenon is well studied and commonly referred to as channel aging \cite{truong_channel_aging, kong_channel_aging, papazafeiropoulos_channel_aging}.

To mitigate the decline in \ac{DL} communication performance caused by channel aging, while avoiding additional communication overhead from increasing the frequency of \ac{CSI} estimation, strategies for channel prediction are investigated.
Different Wiener predictor realizations have been shown to yield estimates that slightly outperform the outdated \ac{CSI} on simulated channels \cite{truong_channel_aging, kong_channel_aging, zheng_channel_aging, papazafeiropoulos_performance_analysis_time_varying_mimo, loeschenbrand_wiener_channel_prediction}.
Recent advancements include improved methods such as Kalman filter \cite{bjoernson_kalman_channel_prediction, kim_kalman_ml_channel_prediction} and machine learning \cite{kim_kalman_ml_channel_prediction, jiang_ml_channel_prediction, wu_ml_channel_prediction}, which have also been evaluated mostly on simulated \ac{CSI}.
Only few authors consider real-world \ac{CSI} for channel prediction, such as \cite{shehzad_ml_channel_prediction, loeschenbrand_ml_channel_prediction}.
All these methods have in common that they are trained based on temporal correlation between consecutive \ac{CSI} samples.
While this may be effective in simple, low-scattering environments, abrupt changes of the \ac{CSI} (e.g., if the \ac{UE} moves behind an obstacle) are barely predictable without spatial information about the environment.

Channel charting \cite{studer_cc} is a self-supervised technique that uses similarity relationships between CSI samples and side information such as timestamps to learn a physically meaningful latent representation of the radio environment, the so-called channel chart.
Due to the spatial consistency of a channel chart, inertia can be leveraged to perform predictive tasks such as \ac{SNR}, handover, and beam prediction \cite{studer_cc, kazemi_cc_snr_prediction, yassine_beam_prediction} through the latent space.
Unlike in \cite{chaaya2024learninglatentwirelessdynamics}, where the prediction of future latent representations has been investigated based on \acp{JEPA} and \ac{UE} velocity information, we learn the channel chart explicitly at first without the knowledge of \ac{UE} velocity, and then predict the \ac{UE}'s position within the channel chart based on the most recent positions.
Additionally, we estimate the future \ac{CSI} based on the predicted channel chart position.

\begin{figure}
    \centering
    \includegraphics[width=0.8\columnwidth]{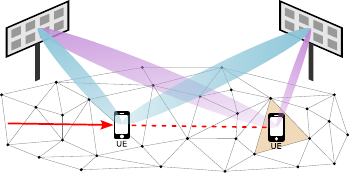}
    \vspace{-0.1cm}
    \caption{Channel charting-based channel prediction: The UE position within the channel chart (black dots) is predicted (dashed red line) from its previous trajectory (solid red line). Delaunay triangulation generates a mesh of triangles among known channel chart positions. The triangle containing the predicted position (orange) is formed by the three samples used for CSI interpolation. Outdated UL beams are visualized in blue, and predicted DL beams in violet.}
    \label{fig:channel_prediction_concept}
\end{figure}

\subsection{Contributions}
We propose a channel prediction method for distributed massive \ac{MIMO} systems that leverages the spatial consistency of a channel chart, as conceptually visualized in Fig. \ref{fig:channel_prediction_concept}.
In particular, multiple distributed massive \ac{MIMO} \ac{BS} antenna arrays independently predict the respective future \ac{CSI} through the latent space.
The array with the best predicted channel is selected for \ac{DL} transmission.
Our method is evaluated on real-world massive \ac{MIMO} measurements from an industrial environment and compared to a Wiener predictor and the outdated \ac{CSI} in terms of achievable sum rate.
The datasets and source code used by this work are publicly available\footnote{\url{https://github.com/phillipstephan/ChannelCharting-ChannelPrediction}}.

\subsection{Outline}
Section \ref{sec:dataset_system_model} describes our dataset and system model.
The problem formulation, array selection strategy and sum rate definition are addressed in Section \ref{sec:channel_prediction}.
After deriving the Wiener predictor in Section \ref{sec:channel_prediction_wiener}, our proposed prediction method is introduced in Section \ref{sec:channel_prediction_cc}.
Finally, Section \ref{sec:results} compares the prediction methods applied to our dataset.
The symbols and notations used throughout this work are shown in Table \ref{tab:notations}.

\begin{table}
    \caption{Symbols and notations used in this paper}\label{tab:notations}
    \vspace{-0.1cm}
    \centering
    \begin{tabular}{r | l}
        $\mathbf A$, $\mathbf b$ & \begin{tabular}{@{}l@{}}Bold letters: Uppercase for matrices and tensors, \vspace{-0.07cm}\\ lowercase for vectors\end{tabular}\\
        $m, N$ & Italic uppercase or lowercase letters: Scalars\\
        $\mathbf{A}^{(l)}$ & Superscript letters: indexing time instant $l$ of tensor $\mathbf{A}$\\
        $\mathbf{A}_{ijk}$ & \begin{tabular}{@{}l@{}}Subscript letters: indexing elements \vspace{-0.07cm}\\ along axes $i,j,k$ of tensor $\mathbf{A}$\end{tabular}\\
        \begin{tabular}{l@{}l@{}l@{}}$\mathbf A_{i::}$ \vspace{-0.07cm}\\$\mathbf A_{ij:}$ \vspace{-0.07cm}\\$\mathbf A_{i:k}$ \end{tabular} & \begin{tabular}{@{}l@{}l@{}}Sub-matrix (and sub-vector) of elements in $i$\textsuperscript{th} entry \vspace{-0.07cm}\\of the first dim. (and $j$\textsuperscript{th} entry of the second dim. \vspace{-0.07cm}\\or $k$\textsuperscript{th} entry of the third dim.) of tensor $\mathbf A$\end{tabular}\\
        $\lVert \mathbf A \rVert_F$, $\lVert \mathbf b \rVert$ & Frobenius (Euclidean) norm of matrix $\mathbf A$ (or vector $\mathbf b$) \vspace{0.02cm}\\
        $m^*$, $\mathbf A^\mathrm{H}$ & Conjugate of scalar $m$ / conj. transpose of matrix $\mathbf A$\\
        $\mathbf a \odot \mathbf b$ & Hadamard (elementwise) product of vectors $\mathbf a$ and $\mathbf b$ \\
    \end{tabular}
\end{table}

\subsection{Limitations}
We present our results on a real-world distributed massive \ac{MIMO} dataset.
The measurement setup is limited to single user scenarios with \ac{UE} velocities that are relatively slow and approximately constant within short time periods.
Furthermore, only one \ac{BS} antenna array is considered for \ac{DL} communication.
For joint \ac{DL} transmission with multiple arrays, phase differences between arrays need to be predicted additionally to prevent destructive interference at the \ac{UE}.

\section{Dataset and System Model}\label{sec:dataset_system_model}

\begin{figure*}
    \centering
    \begin{subfigure}[b]{0.33\textwidth}
        \centering
        \includegraphics[width=0.95\textwidth, trim = 30 200 30 0, clip]{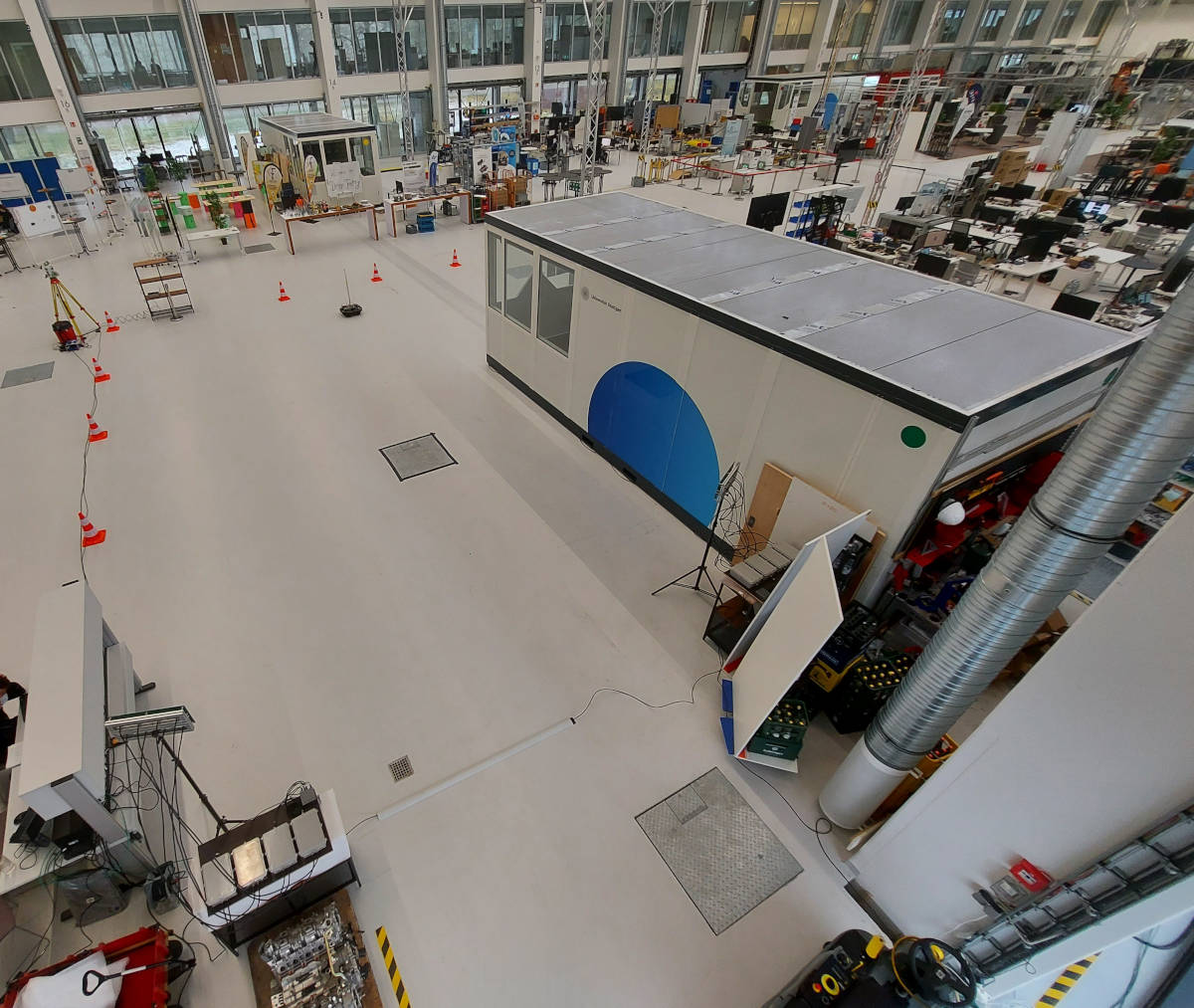}
        \vspace{0.2cm}
        \caption{}
    \end{subfigure}
    \begin{subfigure}[b]{0.32\textwidth}
        \centering
        \begin{tikzpicture}
            \begin{axis}[
                width=0.729\columnwidth,
                height=0.6\columnwidth,
                scale only axis,
                xmin=-15.5,
                xmax=6.1,
                ymin=-18.06,
                ymax=-1.5,
                xlabel = {Coordinate $x_1 ~ [\mathrm{m}]$},
                ylabel = {Coordinate $x_2 ~ [\mathrm{m}]$},
                ylabel shift = -8 pt,
                xlabel shift = -4 pt,
                xtick={-10, -6, -2, 2}
            ]
                \addplot[thick,blue] graphics[xmin=-14.5,ymin=-17.06,xmax=4.1,ymax=-1.5] {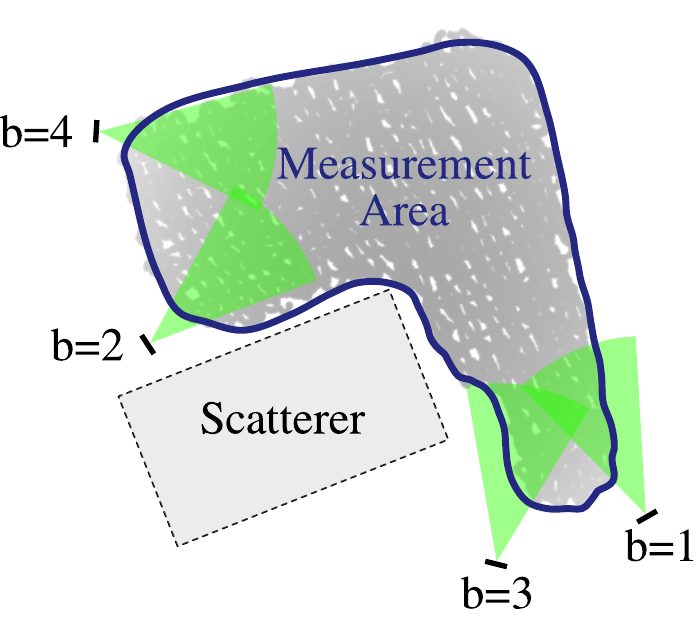};
            \end{axis}
        \end{tikzpicture}
        \vspace{-0.2cm}
        \caption{}
        \label{fig:labelled-area}
    \end{subfigure}
    \begin{subfigure}[b]{0.32\textwidth}
        \centering
        \begin{tikzpicture}
            \begin{axis}[
                width=0.6\columnwidth,
                height=0.6\columnwidth,
                scale only axis,
                xmin=-12.5,
                xmax=2.5,
                ymin=-14.5,
                ymax=-1.5,
                xlabel = {Coordinate $x_1 ~ [\mathrm{m}]$},
                ylabel = {Coordinate $x_2 ~ [\mathrm{m}]$},
                ylabel shift = -8 pt,
                xlabel shift = -4 pt,
                xtick={-10, -6, -2, 2}
            ]
                \addplot[thick,blue] graphics[xmin=-12.5,ymin=-14.5,xmax=2.5,ymax=-1.5] {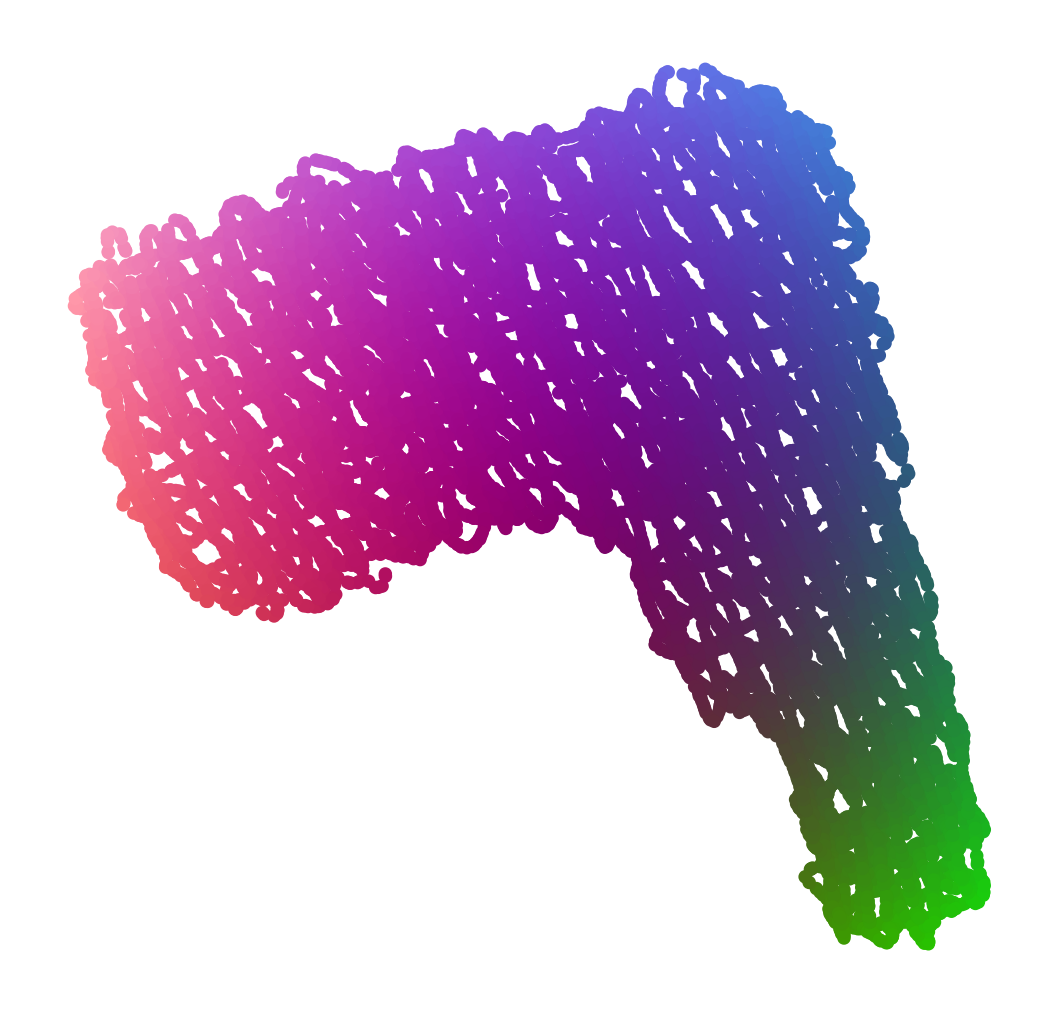};
            \end{axis}
        \end{tikzpicture}
        \vspace{-0.2cm}
        \caption{}
        \label{fig:groundtruth-map}
    \end{subfigure}
    \vspace{-0.3cm}
    \caption{Information about the environment the dataset was measured in: The figure shows (a) a photograph of the environment, (b) a top view map and (c) a scatter plot of colorized ``ground truth'' positions of datapoints in $\mathcal D_\mathrm{train}$, colorized with the measured delay spread. The antenna arrays in the map are drawn to scale as black rectangles and their viewing direction is indicated by the green sectors.}
    \vspace{-0.1cm}
    \label{fig:industrial_environment}
\end{figure*}

Throughout this paper, we use a subset of the \emph{dichasus-cf0x} \cite{dataset-dichasus-cf0x} dataset, which has been captured by our channel sounder \ac{DICHASUS} \cite{dichasus2021} in an industrial environment.
The system is built of a massive \ac{MIMO} \ac{BS} that has $B=4$ distributed uniform planar arrays, each consisting of $M=8$ patch antennas (2 rows, 4 columns), and a \ac{UE} equipped with a single dipole antenna.
All \ac{BS} antennas are synchronized to each other through a reference signal in terms of frequency, time, and phase \cite{geometry_phase_time_sync}.
For the discrete time instances $l = 1, \ldots, L$, frequency-domain channel coefficients $\mathbf{H}^{(l)} \in \mathbb{C}^{B \times M \times N_\mathrm{sub}}$ between the \ac{UE} and all $B\times M$ \ac{BS} antennas and $N_\mathrm{sub} = 1024$ \ac{OFDM} subcarriers are measured and labeled with the corresponding ground truth \ac{UE} position $\mathbf{x}^{(l)} \in \mathbb{R}^2$ and timestamp $t^{(l)} \in \mathbb{R}$, whereas $t^{(l)} - t^{(l-1)} = 0.192 \si{s}$.
The dataset can be formalized as:
\[
    \text{Dataset}: \mathcal D = \left\{ \left(\mathbf H^{(l)}, \mathbf x^{(l)}, t^{(l)} \right) \right\}_{l = 1, \ldots, L}
\]
The training set $\mathcal D_\mathrm{train}$, used for learning the channel chart and Wiener predictor, and the prediction set $\mathcal D_\mathrm{pred}$, on which channel prediction is performed, are both distinctively sampled from the same measurement area in \emph{dichasus-cf02}, \emph{dichasus-cf03} and \emph{dichasus-cf04}.
In the following, we keep the notation from $\mathcal D$ for the training set and denote the prediction set as
\[
    \mathcal D_\mathrm{pred} = \left\{ \left(\mathbf H'^{(l)}, \mathbf x'^{(l)}, t'^{(l)} \right) \right\}_{l = 1, \ldots, L'}
\]
These datasets contain $L=20827$ and $L'=20841$ datapoints, respectively.
The average \ac{UE} velocity is approximately constant at around $0.3\si{\meter\per\second}$.
Note that the ground truth \ac{UE} positions $\mathbf x^{(l)}$ and $\mathbf x'^{(l)}$ are only used for evaluation.

\section{Channel Prediction in Distributed Massive MIMO Systems}\label{sec:channel_prediction}
We consider the aforementioned distributed massive \ac{MIMO} system operating in \ac{TDD} mode, ensuring \ac{UL}-\ac{DL} reciprocity.
\ac{UL} \ac{CSI} is estimated at all \ac{BS} antenna arrays.
Channel prediction is performed to mitigate decreased \ac{DL} throughput caused by outdated \ac{CSI} estimates.
In our scenario, only the array with the best predicted channel is selected for \ac{DL} communication.

\subsection{Channel Prediction}
\ac{CSI} prediction is performed on $\mathcal D_\mathrm{pred}$, which can be seen as a stream of continuously sampled, previously unseen \ac{CSI} samples from a single \ac{UE}.
The prediction problem can be stated as follows: At any time instant $l$, the system aims to predict the future \ac{CSI} $\mathbf{\hat{H}}'^{(l+p)}$ at time instant $l+p$ from a finite memory $\mathbf{\Tilde{H}}'^{(l,K)}$, whereas $p$ is the prediction horizon and $K$ is the size of the memory, which consists of the $K$ most recent \ac{CSI} samples, stacked in a tensor as
\[
    \mathbf{\Tilde{H}}'^{(l,K)} = \hspace{-0.1cm} \left[ \mathbf H'^{(l)}, \mathbf H'^{(l-1)}, \ldots, \mathbf H'^{(l-K+1)} \right] \hspace{-0.1cm} \in \mathbb{C}^{B \times M \times N_\mathrm{sub} \times K}
\]
such that the sum rate for $\mathbf{\hat{H}}'^{(l+p)}$ and the true \ac{CSI} $\mathbf{H}'^{(l+p)}$ is maximized.
The sum rate is computed as defined in Section \ref{sec:sum_rate}.
We compare the following prediction methods:
\begin{itemize}
    \item Outdated \ac{CSI}, i.e., $\mathbf{\hat{H}}_\mathrm{outdated}'^{(l+p)} = \mathbf{H}'^{(l)}$
    \item Wiener predictor
    \item Channel charting-based predictor
\end{itemize}

\subsection{Array Selection Strategy}
The predicted \ac{UL} \ac{CSI} tensor $\mathbf{\hat{H}}'^{(l+p)} \in \mathbb{C}^{B \times M \times N_\mathrm{sub}}$ contains the \ac{CSI} for all $B$ \ac{BS} antenna arrays, whereas only one of those arrays is used for \ac{DL} communication.
For each predicted \ac{CSI} sample at time instant $l+p$, the \ac{BS} estimates the array with the best predicted channel as
\[
    \hat{b}^{(l+p)} = \argmax_b \sum_n^{N_\mathrm{sub}} \left| \left(\mathbf{\hat{H}}_{b:n}'^{(l+p)}\right)^\mathrm{H} \left(\mathbf{\hat{H}}_{b:n}'^{(l+p)}\right) \right|.
\]
This strategy is applied for all considered prediction methods.

\subsection{Sum Rate}\label{sec:sum_rate}
We compute the sum rate based on the received \ac{DL} power achieved with the predicted \ac{CSI}.
Assuming that the transmit power is equally allocated to all subcarriers, the power for each \ac{BS} antenna array $b$ and subcarrier $n$ is computed as
\[
    P_{bn}^{(l+p)} = \frac{\left| \left(\mathbf H_{b:n}'^{(l+p)}\right)^\mathrm{H} \left(\mathbf{\hat{H}}_{b:n}'^{(l+p)}\right) \right|^2}{N_\mathrm{sub}\cdot\left\lVert \mathbf{\hat{H}}_{b:n}'^{(l+p)} \right\rVert^2}.
\]
Assuming a constant noise power $N_0 = \mathbb{E}\left[P / \mu\right]$ at each subcarrier and an average \ac{SNR} of $\mu = 100$, the sum rate at array $b$ is computed as
\[
    \mathrm{SR}_{b}^{(l+p)} = \frac{1}{N_\mathrm{sub}}\sum_{n=1}^{N_\mathrm{sub}} \mathrm{log}_2 \left(1+ \frac{1}{N_0} \cdot P_{bn}^{(l+p)}\right).
\]

\section{Wiener Predictor}\label{sec:channel_prediction_wiener}
We implement the Wiener predictor based on the multi-step predictor introduced in \cite{loeschenbrand_wiener_channel_prediction}.
However, we need to adapt the method to our system setup.
As described in Section \ref{sec:dataset_system_model}, all \ac{BS} antennas are synchronized to each other, but not to the \ac{UE}, causing a global random phase rotation $s^{(l)} = e^{j\theta^{(l)}} \in \mathbb{C}$ with $\lvert s^{(l)} \rvert = 1$ at each \ac{CSI} sample.
The phase shift $\theta^{(l)}$ is modeled as a uniformly distributed random variable $\theta^{(l)} \sim \mathcal{U}\left(0, \, 2\pi\right) \in \mathbb{R}$.
With $l'$ being the time lag, the cross-correlation function between any colocated pair of antennas with indices $m_1$ and $m_2$ at \ac{BS} array $b$ and subcarrier $n$ is given by
\begin{equation}
    \begin{split}
        R_{\mathbf{H}_{bm_1n} \mathbf{H}_{bm_2n}}\left(l'\right) = \mathbb{E}\left[\mathbf{H}_{bm_1n}^{(l+l')} e^{j\theta^{(l+l')}} \mathbf{H}_{bm_2n}^{(l)*} e^{-j\theta^{(l)}} \right]\\
        = \mathbb{E}\left[\mathbf{H}_{bm_1n}^{(l+l')} \mathbf{H}_{bm_2n}^{(l)*}\right] \mathbb{E} \left[ e^{j\left(\theta^{(l+l')}-\theta^{(l)}\right)} \right].     
    \end{split}
\end{equation}
Since $\theta^{(l+l')}$ and $\theta^{(l)}$ are independent,
\[
    \mathbb{E} \left[ e^{j\left(\theta^{(l+l')}-\theta^{(l)}\right)} \right] = 0 \quad \text{for} \quad l' \neq 0,
\]
and therefore,
\[
    R_{\mathbf{H}_{bm_1n} \mathbf{H}_{bm_2n}}\left(l'\right) = 0 \quad \text{for} \quad l' \neq 0.
\]
This implies that a Wiener predictor is not directly applicable to the channel coefficients if they experience a random global phase rotation at each datapoint.
However, the sample autocorrelation matrix of the vector $\mathbf{H}_{b:n}^{(l)}$ containing the channel coefficients for all colocated antenna elements at array $b$, subcarrier $n$ and time instant $l$ is invariant to global phase rotation.
We define the sample autocorrelation matrix of $\mathbf{H}_{b:n}^{(l)}$ as
\[
    \mathbf{Z}_{b::n}^{(l)} = \mathbf{H}_{b:n}^{(l)}\mathbf{H}_{b:n}^{(l)\mathrm{H}} \in \mathbb{C}^{M \times M},
\]
whereas the phase of element $\mathbf{Z}_{bm_1m_2n}^{(l)}$ can be interpreted as the phase difference between the antenna elements $m_1$ and $m_2$ at array $b$ and subcarrier $n$. Since all entries of $\mathbf{Z}^{(l)}$ are invariant to global phase rotation, we propose to apply a Wiener predictor for each entry of $\mathbf{Z}^{(l)}$ separately.
Under the assumption of a wide-sense stationary fading process, the elementwise temporal autocorrelation functions are given by
\[
    \mathbf{R}^{(l')} = \mathbb{E}\left[\mathbf{Z}^{(l+l')} \odot \mathbf{Z}^{(l)*}\right] \in \mathbb{C}^{B \times M \times M \times N_\mathrm{sub}}.
\]
The correlation coefficients for delay $l'$ are then computed as
\[
    r_{bm_1m_2n}^{(l')} = \frac{\mathbf{R}_{bm_1m_2n}^{(l')}}{\mathbf{R}_{bm_1m_2n}^{(0)}} \in \mathbb{C}.
\]

Depending on the memory size $K$ and the delay $p$, the respective correlation coefficients are collected in the vector
\[
    \mathbf{\delta}_{bm_1m_2n,K,p} = \left[r_{bm_1m_2n}^{(p)}, r_{bm_1m_2n}^{(p+1)}, \ldots, r_{bm_1m_2n}^{(p+K-1)}\right]
\]
and the symmetric Toeplitz matrix
\[
    \mathbf{\Delta}_{bm_1m_2n,K,p} \hspace{-0.05cm} = \hspace{-0.1cm}
    \begin{bmatrix}
        r_{bm_1m_2n}^{(p)}, & r_{bm_1m_2n}^{(p+1)}, & \ldots, & r_{bm_1m_2n}^{(p+K-1)} \\
        r_{bm_1m_2n}^{(p+1)}, & r_{bm_1m_2n}^{(p)}, & \ldots, & r_{bm_1m_2n}^{(p+K-2)} \\
        \vdots & \vdots & \ddots & \vdots \\
        r_{bm_1m_2n}^{(p+K-1)}, & r_{bm_1m_2n}^{(p+K-2)}, & \ldots, & r_{bm_1m_2n}^{(p)}
    \end{bmatrix}\hspace{-0.1cm}.
\]
The filter coefficients of the Wiener predictor are given by
\[
    \mathbf{V}_{bm_1m_2n,K,p} = \mathbf{\delta}_{bm_1m_2n,K,p} \mathbf{\Delta}_{bm_1m_2n,K,0}^{-1} \in \mathbb{C}^{K}.
\]
The elements of the predicted sample autocorrelation matrix are then estimated as
\[
    \mathbf{\hat{Z}}_{bm_1m_2n}'^{(l+p)} = \mathbf{V}_{bm_1m_2n,K,p} \left(\mathbf{\Tilde{H}}_{bm_1n:}'^{(l,K)} \odot \mathbf{\Tilde{H}}_{bm_2n:}'^{(l,K) *}\right) \in \mathbb{C}.
\]
Estimating the actual channel coefficients poses the least squares optimization problem:
\begin{equation}\label{eq:sample_autocorrelation_optimization}
    \mathbf{\hat{H}}_{b:n}'^{(l+p)} = \argmin_{\boldsymbol{\vartheta}} \lVert \mathbf{\hat{Z}}_{b::n}'^{(l+p)} - \boldsymbol{\vartheta}\boldsymbol{\vartheta}^\mathrm{H} \lVert_F^2.
\end{equation}
With $\mathbf{\hat{Z}}_{b::n}'^{(l+p)}$ being Hermitian and positive semidefinite, $\mathbf{\hat{H}}_{b:n}'^{(l+p)}$ can be obtained from the set of eigenvectors and corresponding eigenvalues $\{\left(\boldsymbol{\vartheta}_q, \lambda_q\right)\}$ of $\mathbf{\hat{Z}}_{b::n}'^{(l+p)}$, such that
\[
    \mathbf{\hat{H}}_{b:n}'^{(l+p)} = \sqrt{\lambda_{q_\mathrm{princ}}} \frac{\boldsymbol{\vartheta}_{q_\mathrm{princ}}}{\lVert\boldsymbol{\vartheta}_{q_\mathrm{princ}}\rVert}
    \quad \text{with} \quad q_\mathrm{princ} = \argmax_q{\lambda_q}.
\]
Refer to \cite[Sec.\ 4]{geometry_phase_time_sync} for detailed derivation of solving (\ref{eq:sample_autocorrelation_optimization}).

\section{Channel Charting-Based Channel Prediction}\label{sec:channel_prediction_cc}
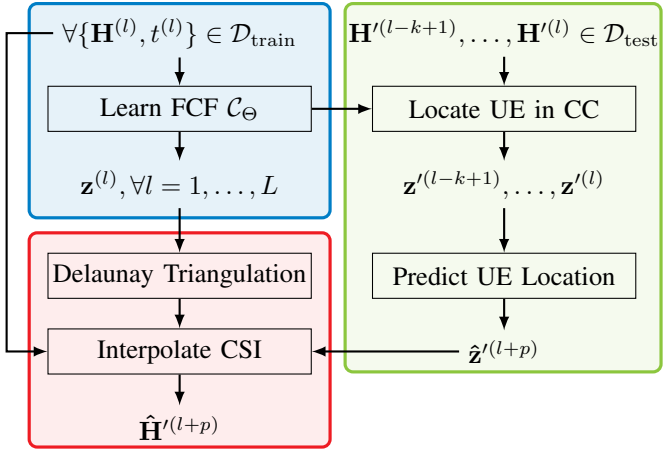
\begin{figure}
    \centering
        \begin{tikzpicture}

    \draw[mittelblau, very thick, rounded corners = 3pt, fill = mittelblau!10!] (-2.0,-1.45)  rectangle (2.0,1.4);
    
    \draw[rot, very thick, rounded corners = 3pt, fill = rot!8!] (-2.0,-1.65)  rectangle (2.0,-4.5);
    
    \draw[apfelgruen, very thick, rounded corners = 3pt, fill = apfelgruen!9!] (2.2,1.4)  rectangle (6.4,-3.5);

    \node (FCF) [align = center, draw, minimum width = 3.5cm, minimum height = 0.6cm, inner sep = 0mm] {Learn \ac{FCF} $\mathcal C_\Theta$};
    \node (Htrain) at ($(FCF.north) + (0, 0.7)$) {$\forall \{\mathbf H^{(l)}, t^{(l)}\} \in \mathcal{D}_\mathrm{train}$};
    
    \node (CC) at ($(FCF.south) - (0, 0.7)$) {$\mathbf{z}^{(l)}, \forall l = 1,\ldots,L$};

    \draw [thick, -latex] (Htrain) -- ($(FCF.north) + (0, 0.0)$);
    \draw [thick, -latex] (FCF.south) -- (CC.north);
    
    \node [draw, minimum width = 3.5cm, minimum height = 0.6cm, inner sep = 0mm] (estccpos) [right = 0.8cm of FCF] {Locate \ac{UE} in CC};

    \node (Htest) at ($(estccpos.north) + (0, 0.7)$) {$\mathbf H'^{(l-k+1)},\ldots,\mathbf H'^{(l)} \in \mathcal{D}_\mathrm{test}$};
    
    \draw [thick, -latex] (Htest) -- ($(estccpos.north) + (0, 0.0)$);
    \draw [thick, -latex] (FCF.east) -- (estccpos.west);
    
    \node (cctest) at ($(estccpos.south) - (0, 0.7)$) {$\mathbf{z}'^{(l-k+1)},\ldots,\mathbf{z}'^{(l)}$};
    
    \draw [thick, -latex] (estccpos.south) -- (cctest.north);
    
    \node [draw, minimum width = 3.5cm, minimum height = 0.6cm, inner sep = 0mm] (predccpos) at ($(cctest.south) - (0, 0.9)$) {Predict \ac{UE} Location};

    \draw [thick, -latex] (cctest.south) -- (predccpos.north);
    
    \node (ccpred) at ($(predccpos.south) - (0, 0.7)$) {$\mathbf{\hat{z}}'^{(l+p)}$};
    
    \draw [thick, -latex] (predccpos.south) -- (ccpred.north);
    
    \node [draw, minimum width = 3.5cm, minimum height = 0.6cm, inner sep = 0mm] (delaunay) at ($(CC.south) - (0, 0.9)$) {Delaunay Triangulation};
    
    \draw [thick, -latex] (CC.south) -- (delaunay.north);
    
    \node [draw, minimum width = 3.5cm, minimum height = 0.6cm, inner sep = 0mm] (interpolate) at ($(delaunay.south) - (0, 0.7)$) {Interpolate \ac{CSI}};
    
    \draw [thick, -latex] (delaunay.south) -- (interpolate.north);
    
    \draw [thick, -latex] (Htrain.west) -| (-2.3,0) |- (interpolate.west);
    
    \draw [thick, -latex] (ccpred.west) -| ($(interpolate.east) + (0.5, 0)$) -- (interpolate.east);
    
    \node (csipred) at ($(interpolate.south) - (0, 0.7)$) {$\mathbf{\hat{H}}'^{(l+p)}$};
    
    \draw [thick, -latex] (interpolate.south) -- (csipred.north);
\end{tikzpicture}
    \vspace{-0.5cm}
    \caption{The three major steps of channel charting-based channel prediction: Learning the \ac{FCF} (red), predicting the \ac{UE}'s location within the channel chart (green), and interpolating between known \ac{CSI} samples (red).}
    \label{fig:method_overview}
\end{figure}

The three major steps of our channel prediction method are visualized in Fig. \ref{fig:method_overview}.
At first, the \ac{FCF} is learned on $\mathcal{D}_\mathrm{train}$ and the estimated channel chart positions are stored at the \ac{BS}.
Then, the channel chart positions on $\mathcal{D}_\mathrm{pred}$ are inferred from the \ac{FCF}.
Inertia within the channel chart allows for predicting future positions from a sequence of previous ones.
In the third step, Delauny triangulation is applied to the channel chart positions of $\mathcal{D}_\mathrm{train}$ to create a mesh of triangles.
If the predicted position from the previous step lies within a triangle, the \ac{CSI} is predicted based on linear interpolation between the triangle’s vertices.
Fig. \ref{fig:channel_prediction_concept} illustrates the concept of this method comprehensibly.
Further details on the individual steps are provided in the following sections.

\subsection{Dissimilarity Metric-Based Channel Charting}\label{sec:dissimilarity_cc}
Our approach for \emph{dissimilarity metric-based channel charting}, as described in \cite{stephan2024angle}, relies on dissimilarities $d_{i,j}$ ("pseudo-distances") between any pair of datapoints with indices $i$ and $j$ in the training set, to learn a channel chart jointly for all arrays.
In particular, the \emph{geodesic, fused} dissimilarity metric \cite{stephan2024angle} is employed, which computes dissimilarities based on the \ac{ADP} at all \ac{BS} antenna arrays and information about timestamp differences.
Learning the \ac{FCF}, which maps the high-dimensional \ac{CSI} to the low-dimensional latent space, i.e., the channel chart, is an optimization problem that minimizes the error between the dissimilarity matrix and the point-to-point distances in the channel chart.
Since no ground truth information is available, the learned channel chart embodies the radio environment in a transformed version of the physical coordinates.
We implement the \ac{FCF} as a \ac{DNN} $\mathcal{C}_\Theta: \mathbf{f}^{(l)} \rightarrow \mathbf{z}^{(l)}$, where $\mathbf{f}^{(l)}$ is a \ac{CSI} feature derived from $\mathbf{H}^{(l)}$ serving as \ac{DNN} input, and $\mathbf{z}^{(l)}$ is the respective estimated channel chart position.
As described in \cite{stephan2024angle}, we use sample autocorrelations of time-domain \ac{CSI} as input features.
The advantage of using a \ac{DNN} to model the \ac{FCF} is its ability to infer channel chart positions for new datapoints without retraining.
During training, the \ac{DNN} $\mathcal{C}_\Theta$ is embedded in a Siamese network, and the related Siamese loss function is applied as
\begin{equation}
\mathcal{L}_\mathrm{Siamese}=\sum\nolimits_{i=1}^{L-1}\sum\nolimits_{j=i+1}^L \frac{\left(d_{i, j}-\Vert\mathbf{z}^{(i)}-\mathbf{z}^{(j)}\Vert_2\right)^2}{d_{i, j} + \beta},
\label{eq:siameseloss}
\end{equation}
with $\beta$ being a hyperparameter to weight either the absolute squared error or the normalized squared error higher.
The channel chart positions $\{\mathbf{z}^{(l)}\}_{l=1}^L$ are linked to their corresponding \ac{CSI} tensors $\{\mathbf{H}^{(l)}\}_{l=1}^L$ and stored at the \ac{BS}.

\subsection{Predict User Location within Channel Chart}
The position of a \ac{UE} within the channel chart, while traveling along the trajectories of $\mathcal{D}_\mathrm{pred}$, can be estimated from the learned \ac{FCF} and the \ac{CSI} features as $\mathbf{z}'^{(l)} = \mathcal{C}_\Theta \left(\mathbf{f}'^{(l)}\right)$.
The spatial consistency of the channel chart and the concept of inertia allow for predicting the future channel chart position $\mathbf{\hat{z}}'^{(l+p)}$ by linearly extrapolating the $K$ most recent estimates $\mathbf{z}'^{(l-K+1)},\ldots,\mathbf{z}'^{(l)}$.
Although a more sophisticated approach using \acp{JEPA} and knowledge of the \ac{UE} velocity has been proposed in \cite{chaaya2024learninglatentwirelessdynamics}, we rely on linear extrapolation, since velocity information is not available in our setup, and optimizing for latent space prediction exceeds the scope of this work.

\subsection{Linear Interpolation-Based CSI-Predictor (``CC-interp'')}
At first, the Delaunay triangulation is computed for the channel chart positions in $\mathcal{D}_\mathrm{train}$ to obtain a mesh of triangles between all datapoints.
If the predicted channel chart position $\mathbf{\hat{z}}'^{(l+p)}$ lies within a triangle, the barycentric coordinate vector $\mathbf{c} \in \mathbb{R}^3$ with $\sum_i \mathbf{c} = 1$ is derived from the relative distances to the three known channel chart positions $\mathbf{z}^{(\Delta 1)}, \mathbf{z}^{(\Delta 2)}, \mathbf{z}^{(\Delta 3)}$ forming the triangle.
Linear barycentric interpolation between the respective \ac{CSI} samples at base station $b$ and subcarrier $n$, given the random global phase rotation $e^{j \theta^{(l)}}$ at each datapoint, can be expressed by the optimization problem
\[
    \left(\mathbf{\hat{H}}_{b:n}'^{(l+p)}, \boldsymbol{\theta}\right) = \argmin_{\left(\boldsymbol{\vartheta}, \boldsymbol{\theta}\right)} \sum_{i=1}^3 c_i \lVert \mathbf{H}_{b:n}^{(\Delta i)} - e^{j \theta^{(\Delta i)}}\boldsymbol{\vartheta} \rVert^2.
\]
Similarly to Section \ref{sec:channel_prediction_wiener}, we can eliminate the dependency on $\boldsymbol{\theta}$ by taking sample autocorrelations into account.
The optimization problem can then be reformulated to
\begin{equation}\label{eq:sample_autocorrelation_optimization_2}
    \mathbf{\hat{H}}_{b:n}'^{(l+p)} = \argmin_{\boldsymbol{\vartheta}} \lVert \mathbf{\hat{Z}}_{b::n}'^{(l+p)} - \boldsymbol{\vartheta}\boldsymbol{\vartheta}^\mathrm{H} \rVert_F^2,
\end{equation}
with $\mathbf{\hat{Z}}_{b::n}'^{(l+p)}$ being estimated as
\[
    \mathbf{\hat{Z}}_{b::n}'^{(l+p)} = \frac{1}{3} \sum_{i=1}^3 c_i \left( \mathbf{H}_{b:n}^{(\Delta i)} \mathbf{H}_{b:n}^{(\Delta i) \mathrm{H}} \right).
\]
Analogous to Eq. \ref{eq:sample_autocorrelation_optimization} in Section \ref{sec:channel_prediction_wiener}, $\mathbf{\hat{H}}_{b:n}'^{(l+p)}$ is derived from the set of eigenvectors and corresponding eigenvalues $\{\left(\boldsymbol{\vartheta}_q, \lambda_q\right)\}$ of $\mathbf{\hat{Z}}_{b::n}'^{(l+p)}$, such that
\[
    \mathbf{\hat{H}}_{b:n}'^{(l+p)} = \sqrt{\lambda_{q_\mathrm{princ}}} \frac{\boldsymbol{\vartheta}_{q_\mathrm{princ}}}{\lVert\boldsymbol{\vartheta}_{q_\mathrm{princ}}\rVert}
    \quad \text{with} \quad q_\mathrm{princ} = \argmax_q{\lambda_q}.
\]

\subsection{CSI-Predictor Based on Nearest Neighbor (``CC-NN'')}
A simple alternative method is to predict the \ac{CSI} as the nearest known neighbor in the training set, regarding their corresponding channel chart positions.
In particular, the predicted \ac{CSI} tensor is determined as
\[
    \mathbf{\hat{H}}'^{(l+p)} = \mathbf{H}^{(i_\mathrm{NN})}
\]
with $i_\mathrm{NN}$ being the index for which the distance between the predicted channel chart position $\mathbf{\hat{z}}'^{(l+p)}$ and any channel chart position $\mathbf{z}^{(i)}$ in $\mathcal{D}_\mathrm{train}$ is minimized:
\[
    i_\mathrm{NN} = \argmin_i \lVert \mathbf{z}^{(i)} - \mathbf{\hat{z}}'^{(l+p)} \rVert.
\]
In practice, this method is applicable if the predicted channel chart position does not lie within any triangle formed by known positions in $\mathcal{D}_\mathrm{train}$, making interpolation meaningless.

\section{Results}\label{sec:results}
We evaluate the quality of the channel chart, and the channel prediction performance of our method compared to the Wiener predictor and the outdated \ac{CSI}.
For evaluation of the channel prediction performance, we exclude datapoints where the predicted channel chart position does not lie within a triangle of known channel chart positions.
All predictive tasks are performed with fixed memory size of $K = 25$ samples on a subset of $N_\mathrm{sub}' = 32$ subcarriers that are equally spaced over the whole bandwidth to reduce the computation time.

\subsection{Evaluation of the Channel Chart}
The quality of a channel chart is typically evaluated using performance metrics from dimensionality reduction.
The metrics \ac{CT} and \ac{TW} \cite{trustworthiness_continuity} measure the preservation of local neighborhood dependencies in the channel chart, ranging from $0$ to $1$, whith higher values indicating better performance.
\ac{KS} \cite{kruskal_stress} evaluates the global structure, also ranging from $0$ to $1$, but lower values being preferred.
Additionally, the \ac{MAE} can be computed after applying an optimal affine transformation to the channel chart positions \cite{fraunhofer_cc}.

\begin{figure*}
    \centering
    \begin{subfigure}[b]{0.32\textwidth}
        \centering
    \begin{tikzpicture}
        \begin{axis}[
            width=0.45\columnwidth,
            height=0.45\columnwidth,
            scale only axis,
            xmin=-300,
            xmax=300,
            ymin=-400,
            ymax=400,
            xlabel = {Latent variable $z_1$},
            ylabel = {Latent variable $z_2$},
            ylabel shift = -8 pt,
            xlabel shift = -4 pt,
        ]
            \addplot[thick,blue] graphics[xmin=-300,ymin=-400,xmax=300,ymax=400] {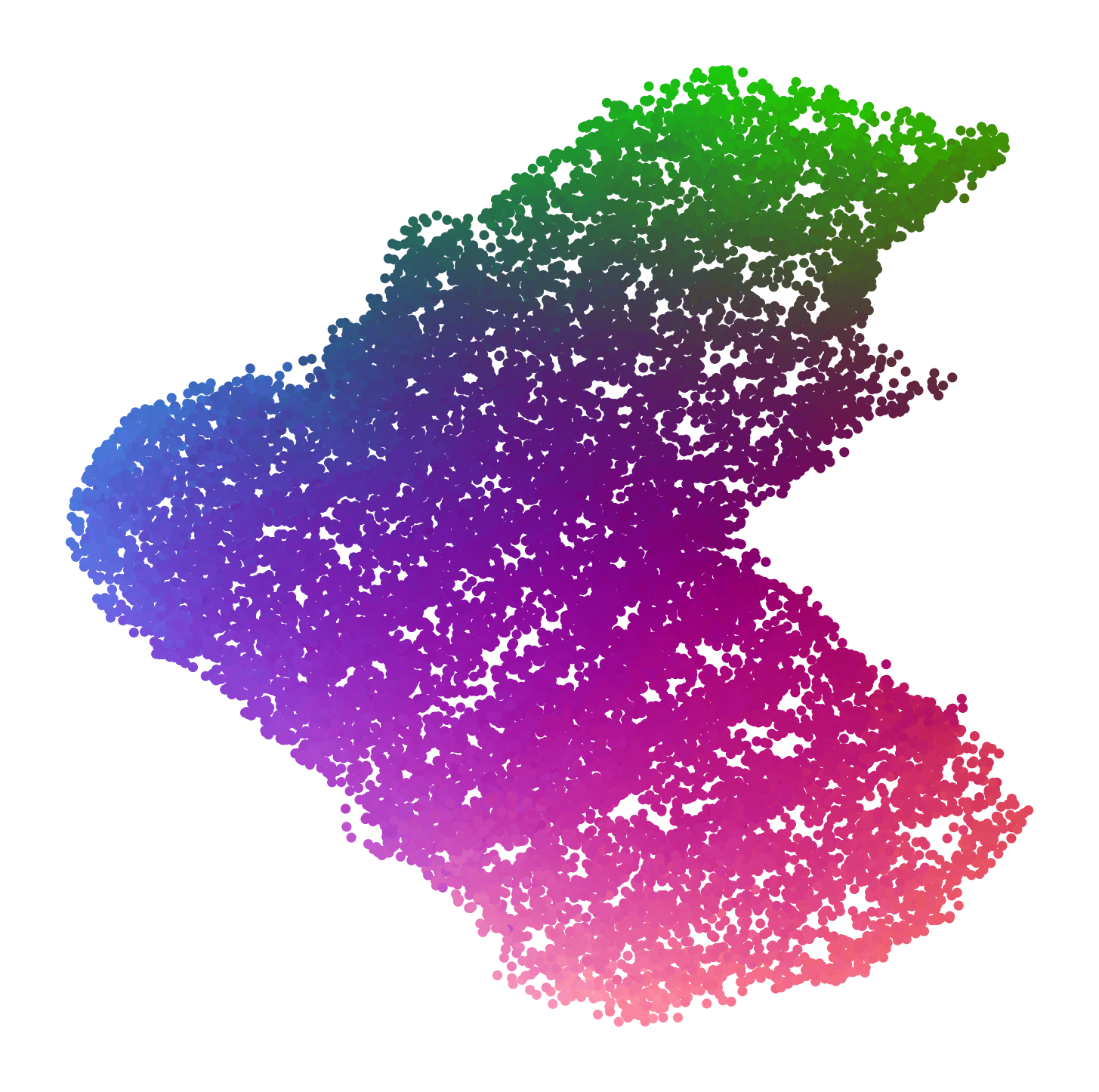};
        \end{axis}
    \end{tikzpicture}
        \vspace{-0.2cm}
        \caption{}
        \label{fig:channel_chart_test}
    \end{subfigure}
    \begin{subfigure}[b]{0.32\textwidth}
        \centering
    \begin{tikzpicture}
        \begin{axis}[
            width=0.45\columnwidth,
            height=0.45\columnwidth,
            scale only axis,
            xmin=-300,
            xmax=300,
            ymin=-400,
            ymax=400,
            xlabel = {Latent variable $z_1$},
            ylabel = {Latent variable $z_2$},
            ylabel shift = -8 pt,
            xlabel shift = -4 pt,
            colormap name = viridis,
            colorbar,
            point meta min=0,
            point meta max=100,
            colorbar style={
                width=0.25cm,
                xshift=-0.25cm,
                ylabel={$\mathrm{MAE}_\mathrm{latent}$},
                ylabel style = {
                    yshift=0.15cm
                }
            }
        ]
            \addplot[thick,blue] graphics[xmin=-300,ymin=-400,xmax=300,ymax=400] {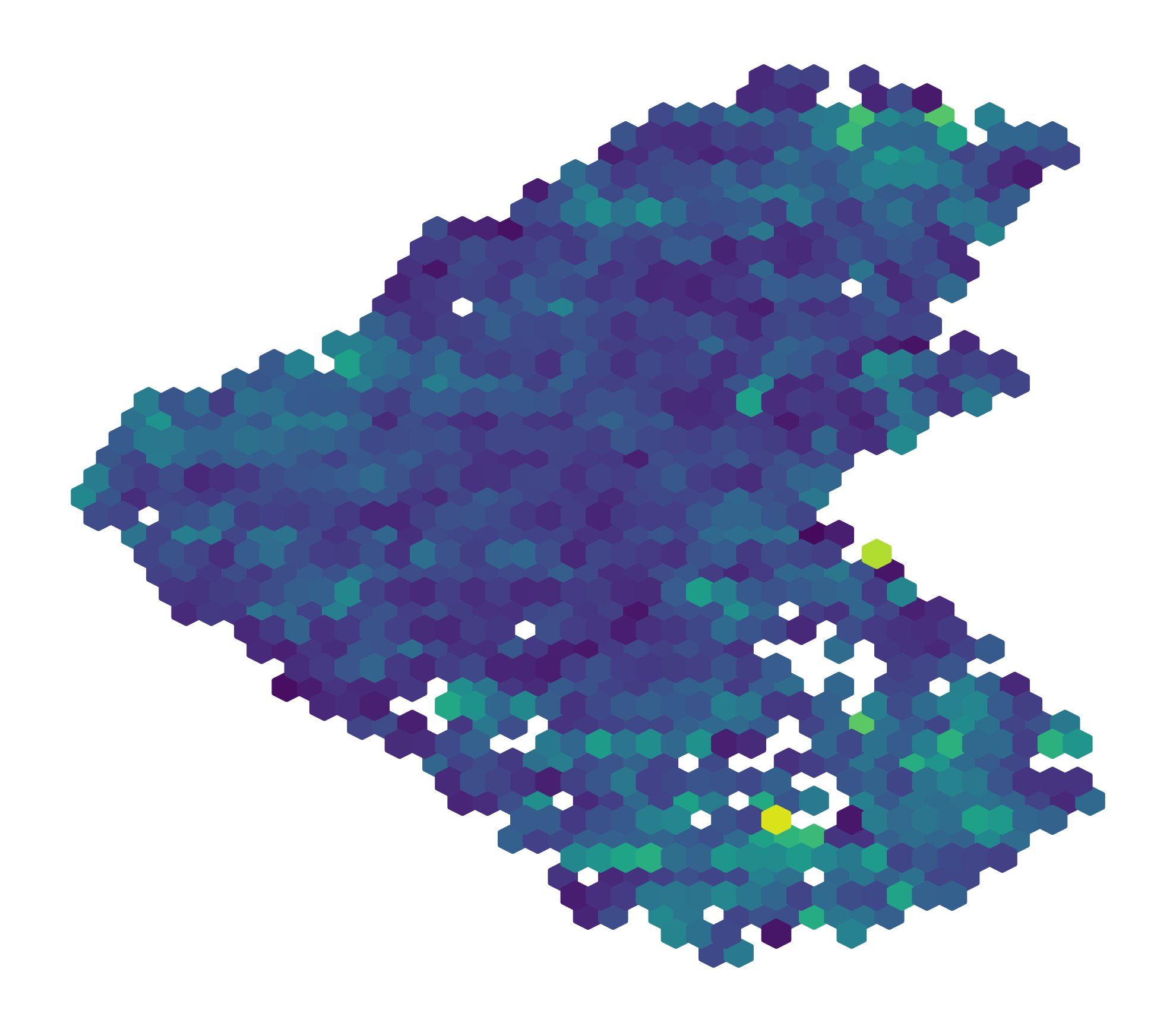};
        \end{axis}
    \end{tikzpicture}
        \vspace{-0.2cm}
        \caption{}
        \label{fig:latent_space_prediction_errors}
    \end{subfigure}
    \begin{subfigure}[b]{0.32\textwidth}
        \centering
    \begin{tikzpicture}
        \begin{axis}[
            width=0.45\columnwidth,
            height=0.45\columnwidth,
            scale only axis,
            xmin=-300,
            xmax=300,
            ymin=-400,
            ymax=400,
            xlabel = {Latent variable $z_1$},
            ylabel = {Latent variable $z_2$},
            ylabel shift = -8 pt,
            xlabel shift = -4 pt,
            colormap name = viridis,
            colorbar,
            point meta min=0,
            point meta max=6.3,
            colorbar style={
                width=0.25cm,
                xshift=-0.25cm,
                ylabel={$\mathrm{SR} \left[\si{bit\per s\per \hertz}\right]$}}
        ]
            \addplot[thick,blue] graphics[xmin=-300,ymin=-400,xmax=300,ymax=400] {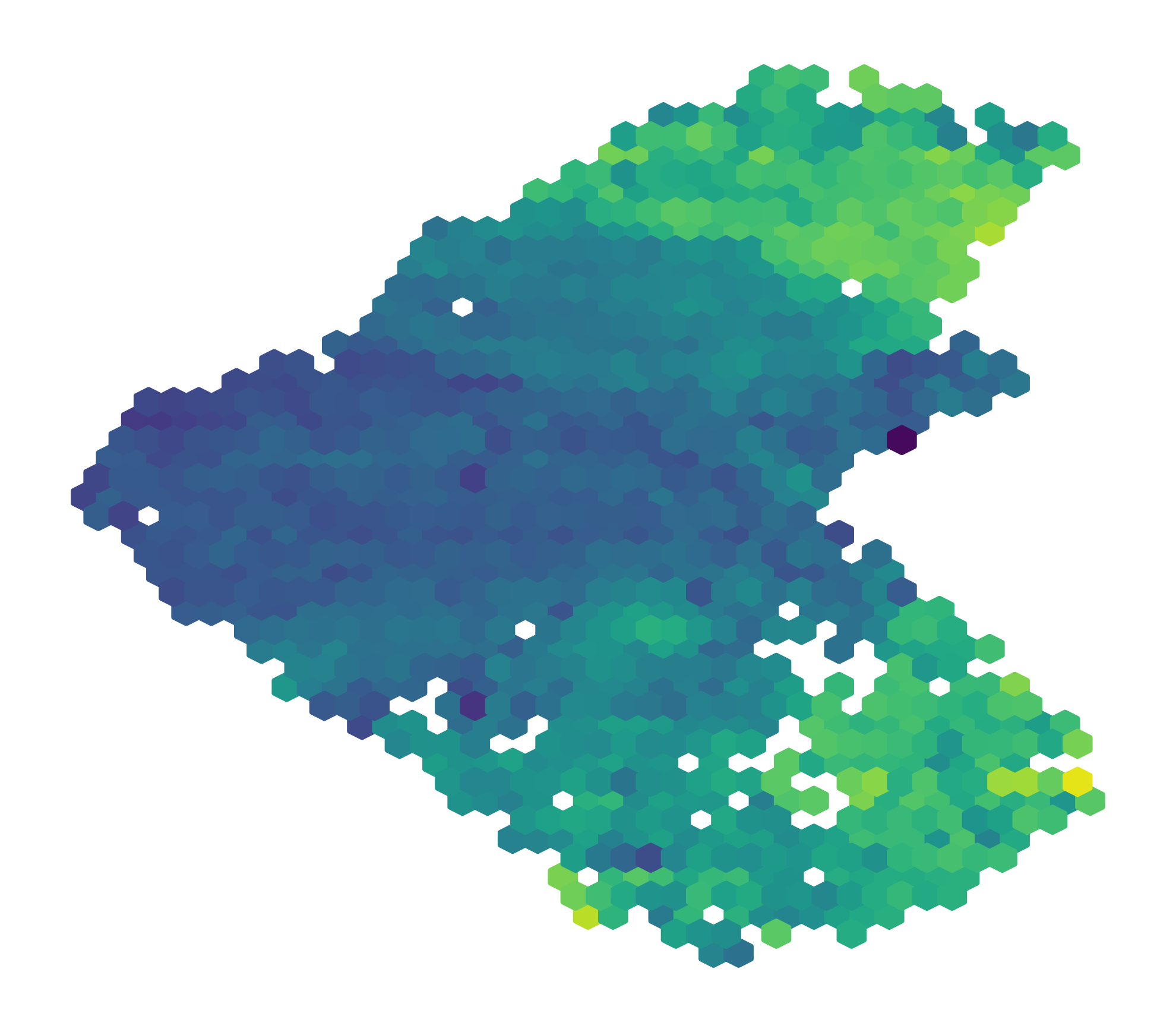};
        \end{axis}
    \end{tikzpicture}
        \vspace{-0.2cm}
        \caption{}
        \label{fig:sum_rate_per_datapoint}
    \end{subfigure}
    \vspace{-0.3cm}
    \caption{Visual evaluation: The figure shows (a) the channel chart positions with preserved coloring from Fig. \ref{fig:groundtruth-map}, (b) the absolute latent space prediction error over the channel chart positions, and (c) the sum rate achieved with the predicted CSI from CC-interp over the channel chart positions.}
    \vspace{-0.1cm}
    \label{fig:visual_evaluation}
\end{figure*}

\begin{table}\label{tab:cc_performance}
	\caption{Evaluation of the Channel Charting Performance.}
	\vspace*{-0.1cm}
	\centering
	\begin{tabular}{ccccc}
	\toprule
 	Dataset & \ac{CT} $\uparrow$ & \ac{TW} $\uparrow$ & \ac{KS} $\downarrow$ & \ac{MAE} $\downarrow$\\
 	\midrule
 	$\mathcal{D}_\mathrm{train}$ & $0.997$ & $0.997$ & $0.082$ & $0.484$ \\
 	$\mathcal{D}_\mathrm{pred}$ & $0.996$ & $0.996$ & $0.083$ & $0.484$ \\
 	\bottomrule
	\end{tabular}
	\label{tab:cc_evaluation}
	\vspace*{-0.3cm}
\end{table}

Fig. \ref{fig:channel_chart_test} depicts a scatter plot of the channel chart positions with preserved coloring from the respective ground truth positions, showing that local neighborhood dependencies are well preserved.
The global structure is also captured, though with a random coordinate transformation.
Table \ref{tab:cc_evaluation} objectively supports this observation, showing the channel charting performance on $\mathcal{D}_\mathrm{train}$ in the first row, and for the inferred positions on $\mathcal{D}_\mathrm{pred}$ in the second row.

\subsection{Latent Space Prediction Error}
The absolute latent space prediction error $\mathrm{MAE}_\mathrm{latent}^{(l+p)} = \lVert \mathbf{\hat{z}}'^{(l+p)} - \mathbf{z}'^{(l+p)} \rVert$ is visualized in Fig. \ref{fig:latent_space_prediction_errors} over all true predicted channel chart positions $\mathbf{z}'^{(l+p)}$ with $l=1,\ldots,L'$ and $p = 15$.
It is clearly visible that the prediction error is relatively small over the whole area compared to the size of the channel chart. The error appears to be slightly larger at areas near the \ac{BS} arrays, which can be explained by the scatterer in the inner corner blocking the \ac{LoS} path for the respective two other arrays.

\subsection{Channel Prediction Performance}
\begin{figure}
    \centering
\begin{tikzpicture}

\definecolor{crimson2143940}{RGB}{214,39,40}
\definecolor{darkgray176}{RGB}{176,176,176}
\definecolor{darkorange25512714}{RGB}{255,127,14}
\definecolor{forestgreen4416044}{RGB}{44,160,44}
\definecolor{lightgray204}{RGB}{204,204,204}
\definecolor{steelblue31119180}{RGB}{31,119,180}

\begin{axis}[
legend cell align={left},
legend style={fill opacity=0.8, draw opacity=1, text opacity=1, draw=lightgray204},
tick align=outside,
width=0.97\columnwidth,height=0.66\columnwidth,
tick pos=left,
x grid style={darkgray176},
xlabel={prediction horizon $p \left[\times 0.192 \si{s}\right]$},
xmajorgrids,
xmin=-0.75, xmax=25.75,
xtick style={color=black},
y grid style={darkgray176},
ylabel={$\mathrm{SR} \left[\si{bit\per s\per \hertz}\right]$},
ymajorgrids,
ymin=2.2240266919136, ymax=3.74013828039169,
ytick style={color=black}
]
\addplot [thick, steelblue31119180, mark=*, mark size=1, mark options={solid}]
table {%
0 3.67122411727905
1 3.44074487686157
2 3.19840431213379
3 3.13575625419617
4 3.09223127365112
5 3.02735996246338
6 2.94504141807556
7 2.88767528533936
8 2.84085917472839
9 2.7890453338623
10 2.73319029808044
11 2.68774580955505
12 2.64862060546875
13 2.60297894477844
14 2.55547523498535
15 2.52450585365295
16 2.49654960632324
17 2.46697306632996
18 2.43386125564575
19 2.4059727191925
20 2.38685321807861
21 2.36749625205994
22 2.35076284408569
23 2.33472466468811
24 2.31659531593323
25 2.29294085502625
};
\addlegendentry{Outdated CSI}
\addplot [thick, darkorange25512714, mark=*, mark size=1, mark options={solid}]
table {%
0 3.19785070419312
1 3.16077303886414
2 3.12855386734009
3 3.0996515750885
4 3.06930541992188
5 3.04509234428406
6 3.02428412437439
7 3.00656890869141
8 2.98530650138855
9 2.96345067024231
10 2.93962812423706
11 2.91777324676514
12 2.89663290977478
13 2.88057494163513
14 2.85666537284851
15 2.84043216705322
16 2.81809687614441
17 2.79461050033569
18 2.77605414390564
19 2.75432300567627
20 2.73432993888855
21 2.7122597694397
22 2.69663953781128
23 2.68076920509338
24 2.6591100692749
25 2.64009737968445
};
\addlegendentry{CC-interp}
\addplot [thick, forestgreen4416044, mark=*, mark size=1, mark options={solid}]
table {%
0 3.0575168132782
1 3.02582287788391
2 2.98953819274902
3 2.96795034408569
4 2.93808603286743
5 2.91016840934753
6 2.89677309989929
7 2.87430143356323
8 2.85608720779419
9 2.83797812461853
10 2.81162285804749
11 2.78975319862366
12 2.76828694343567
13 2.75395274162292
14 2.72989225387573
15 2.70675349235535
16 2.69026637077332
17 2.67532229423523
18 2.65765786170959
19 2.63727617263794
20 2.6180624961853
21 2.5973334312439
22 2.57823848724365
23 2.56173396110535
24 2.54594278335571
25 2.53088021278381
};
\addlegendentry{CC-NN}
\addplot [thick, crimson2143940, mark=*, mark size=1, mark options={solid}]
table {%
0 3.67122411727905
1 3.47204494476318
2 3.29477477073669
3 3.22425436973572
4 3.16731858253479
5 3.10109758377075
6 3.03543400764465
7 2.98575735092163
8 2.94272923469543
9 2.89914631843567
10 2.85786747932434
11 2.82409477233887
12 2.79197812080383
13 2.75918316841125
14 2.72738170623779
15 2.70378637313843
16 2.67594265937805
17 2.65328097343445
18 2.62701416015625
19 2.60377264022827
20 2.57702875137329
21 2.55037045478821
22 2.52540993690491
23 2.49947738647461
24 2.47048139572144
25 2.44094443321228
};
\addlegendentry{Wiener}
\end{axis}

\end{tikzpicture}
    \vspace{-0.6cm}
    \caption{Average sum rate achieved with the \ac{CSI} obtained by different channel prediction methods depending on the prediction horizon $p$.}
    \label{fig:sum_rate}
\end{figure}
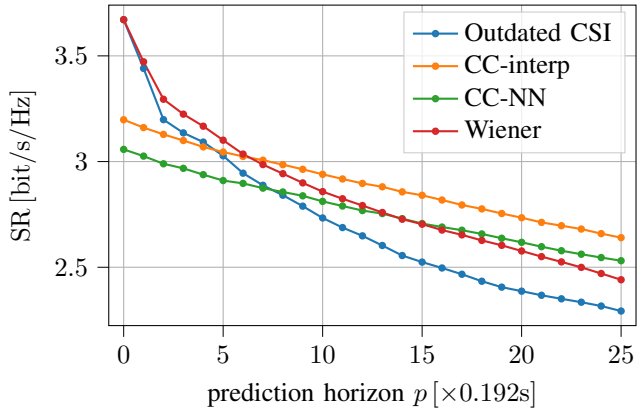

Fig. \ref{fig:sum_rate} shows the average sum rate achieved with the \ac{CSI} obtained by the different prediction methods depending on the prediction horizon $p$, given the array selection strategy from Section \ref{sec:channel_prediction}.
The outdated \ac{CSI} experiences a steep performance decrease for growing prediction horizons.
The Wiener predictor mitigates this effect slightly, whereas its impact increases with $p$.
The advantage of the channel charting-based methods is revealed for larger prediction horizons, though both approaches underperform the outdated \ac{CSI} for small values of $p$.
While the CC-NN method outperforms the outdated \ac{CSI} for $p > 7$ and the Wiener predictor for $p > 13$, the CC-interp method already yields better results for $p > 4$ and $p > 6$, respectively.
With an interval of $0.192\si{\second}$ between consecutive samples, the proposed CC-interp is the best choice for prediction horizons larger than $1 \si{\second}$.
Given the slow \ac{UE} velocity of $0.3 \si{\meter\per\second}$, we expect our method to scale proportionally to higher \ac{UE} velocities on smaller prediction horizons.
Fig. \ref{fig:sum_rate_per_datapoint} shows the sum rate achieved with CC-interp, which is generally higher if the \ac{UE} is near a \ac{BS} array, highlighting the effectiveness of our array selection strategy.

\section{Conclusion and Outlook}
We demonstrated on real-world distributed massive \ac{MIMO} measurements that channel prediction can be performed based on a channel chart.
The \ac{CSI} predicted with our method is more resilient over time than the Wiener predictor and the outdated \ac{CSI}, although the Wiener predictor performs better for short prediction horizons.
However, channel charting is still a relatively new research area with a lot of potential for algorithmic improvements, which will likely enhance the prediction accuracy.
Further, latent space prediction holds potential for optimization, since linear extrapolation of channel chart positions is an idealized assumption that is less effective for curved trajectories.
In addition, more advanced \ac{CSI} interpolation methods may be investigated to improve the prediction performance.
Future work could extend our approach to multi-user scenarios and a joint \ac{DL} communication scheme involving multiple \ac{BS} arrays by additionally taking phase differences between arrays into account.

\bibliographystyle{IEEEtran}
\bibliography{IEEEabrv,references}

\end{document}